%% file: main.tex
\documentclass[12pt]{article}
\usepackage[utf8]{inputenc}
\usepackage{graphicx}
\usepackage{amsmath}
\usepackage{amssymb}
\usepackage{amsthm}
\usepackage{listings}
\usepackage{xcolor}
\usepackage{algorithm2e}
\usepackage{hyperref}

\usepackage{tikz}
\usetikzlibrary{fadings}
\usetikzlibrary{patterns}
\usetikzlibrary{shadows.blur}
\usetikzlibrary{shapes}

\usepackage{caption}
\usepackage{subcaption}

\usepackage{courier}

\definecolor{cppColorBackground}{rgb}{1.,1.,1.}
\definecolor{cppColorComment}{rgb}{0.0,0.47,.8}
\definecolor{cppColorLine}{rgb}{0.6,0.6,0.6}
\definecolor{cppColorString}{rgb}{0,0.501,145}
\definecolor{cppColorKey}{rgb}{0.8,0.5,0}
\definecolor{cppColorDigit}{rgb}{0,0,0.5}
\author{David Algis\\Poitiers University\\Studio Nyx
    \and Berenger Bramas\\INRIA Nancy
    \and Emmanuelle Darles\\Poitiers University
    \and Lilian Aveneau\\Poitiers University
}

\title{Efficient GPU Implementation of Particle Interactions with Cutoff Radius and Few Particles per Cell.}

\begin{document}

\maketitle

\begin{abstract}
    This paper presents novel approaches to parallelizing particle interactions on a GPU when there are few particles per cell and the interactions are limited by a cutoff distance. 
    The paper surveys classical algorithms and then introduces two alternatives that aim to utilize shared memory. 
    The first approach copies the particles of a sub-box, while the second approach loads particles in a pencil along the X-axis. 
    The different implementations are compared on three GPU models using Cuda and Hip.
    The results show that the X-pencil approach can provide a significant speedup but only in very specific cases.
\end{abstract}

%%%%%%%%%%%%%%%%%%%%%%%%%%%%%%%%%%%%%%%%%%%%%%%%%%%%%%%%%%%%%%%%%%%%%%%%%%%%
%%%%%%%%%%%%%%%%%%%%%%%%%%%%%%%%%%%%%%%%%%%%%%%%%%%%%%%%%%%%%%%%%%%%%%%%%%%%
\section{Introduction}
\label{sec:introduction}

Computing pairwise interactions in N-body simulations is a fundamental problem in many scientific and engineering applications.
While the worst case complexity is quadratic relatively to the number of particles, if the interaction kernel decays exponentially with the distance, we can use a cutoff to reduce significantly the number of 
interactions that need to be computed while maintaining a satisfactory level of accuracy.
A classical implementation of this strategy is to rely on a grid-based approach, 
where the simulation domain is divided into a regular grid, and particles are assigned to the grid cells. 
The interactions are then computed between particles in the same cell and between particles in neighboring cells.

In this paper, we focus on the problem of computing pairwise interactions with a cutoff in a grid on a GPU.
More precisely, we consider the case where there are few particles per cell, i.e., the number of particles is such that several cells can be loaded in shared memory, which is a common scenario in many applications, with the aim to develop an efficient implementation using shared memory.
~\footnote{The source code is open-source and available at \url{https://gitlab.inria.fr/bramas/gpu-particle-interaction}} 

The main contribution of this paper is as follows:
\begin{itemize}
\item We propose two approaches to perform the computation while efficiently using shared memory: full load and pencil load. The full load places all the cells of a 3D sub-box into shared memory, whereas the pencil load covers a sub-box with a size greater than one only in the X dimension;
\item We compare both approaches and study their performance and limitations on both NVIDIA and AMD architectures.
\end{itemize}

We also describe a straightforward implementation of the prefix sum algorithm on GPU with minimal memory accesses.
It can be seen as an improvement of the Blelloch's algorithm~\cite{blelloch1989scans, blelloch1990prefix} or a variation of Ladner et al. algorithm~\cite{10.1145/322217.322232}.\footnote{The original Blelloch implementation perform a swap that can be avoided, but similar technics already exist in some implementations,
such as the HIP prefix sum~\url{https://github.com/ROCm/HIP-Examples/blob/master/HIP-Examples-Applications/PrefixSum/PrefixSum.cpp}}

The rest of the paper is organized as follows.
Section~\ref{sec:background} provides background information on the problem of computing pairwise interactions with a cutoff in a grid.
Section~\ref{sec:problem-statement} presents the motivation and problem statement.
Section~\ref{sec:proposed-solutions} describes the proposed solution in detail.
Section~\ref{sec:related-work} discusses the related work.
Section~\ref{sec:experiments} presents the experimental results.
Finally, Section~\ref{sec:conclusion} concludes the paper and outlines directions for future work.

%%%%%%%%%%%%%%%%%%%%%%%%%%%%%%%%%%%%%%%%%%%%%%%%%%%%%%%%%%%%%%%%%%%%%%%%%%%%
%%%%%%%%%%%%%%%%%%%%%%%%%%%%%%%%%%%%%%%%%%%%%%%%%%%%%%%%%%%%%%%%%%%%%%%%%%%%
\section{Background}
\label{sec:background}

The desired behavior of the pairwise interactions between $N$ particles consists in computing the interactions between all pairs of particles $(i, j)$
such that $i \neq j$ and $r_{ij} < r_c$, where $r_{ij}$ is the distance between particles $i$ and $j$, and $r_c$ is the cutoff radius.
The interactions are typically computed using a kernel function $K(r_{ij})$ that depends on the distance between the particles.
The kernel function is usually chosen to be a smooth function that decays rapidly with the distance, such as the Gaussian function or the B-Spline function.
If we perform the computation naively, the complexity of computing all pairwise interactions is $O(N^2)$, which is prohibitive for large values of $N$.
This is why a more advanced implementation consists in using a grid-based approach, where the simulation domain is divided into a regular grid,
and particles are assigned to the grid cells.
The interactions are then computed between particles from the same cell and between particles in neighboring cells, which leads to a linear complexity $O(N)$ when the number of particles per cell is bounded by a constant.

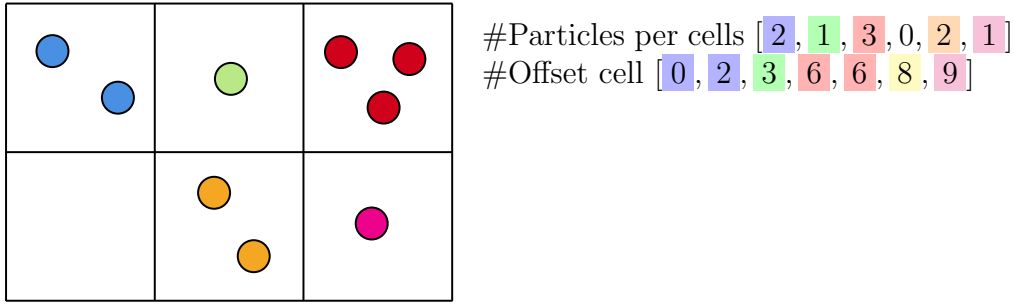
\begin{figure}
	\centering
	\input{tikzPicture/GridExample.tikz}
	\caption{Example of a 2x3 cell grid with 9 particles. We use an array that contains the number of particles per cell and use it to compute the prefix sum.}
    \label{fig:basic-example}
\end{figure}

On GPU, this approach is usually implemented as follows.
The particles' values, such as their positions, are stored in a structure-of-arrays (SoA) format, where each component of the position vector is stored in a separate array.
For each particle, we compute its cell index using its physical position, without moving the particles in memory.
This operation is done in parallel.
Then, we count the number of particles per cell, which can also be done in parallel by using atomic operations.
The resulting array is then used to compute in parallel the prefix sum, which allows knowing where the particles that belong to a given cell should be located in the arrays (Figure \ref{fig:basic-example}).
We use the prefix array to move the particles in a secondary array (this operation is not done in-place), which can be done in parallel using again atomic operations to select the positions of the particles in their corresponding cell's memory chunk.
Finally, the interactions are computed between particles in the same cell and between particles in neighboring cells and their positions are updated.

%\begin{figure}
 %   \centering
  %  \includegraphics[width=\textwidth, height=0.3\textheight, keepaspectratio]{./Figures/page_01}
   % \caption{}
    %\label{fig:basic-example}
%\end{figure}

%%%%%%%%%%%%%%%%%%%%%%%%%%%%%%%%%%%%%%%%%%%%%%%%%%%%%%%%%%%%%%%%%%%%%%%%%%%%
%%%%%%%%%%%%%%%%%%%%%%%%%%%%%%%%%%%%%%%%%%%%%%%%%%%%%%%%%%%%%%%%%%%%%%%%%%%%
\section{Motivation and Problem Statement}
\label{sec:problem-statement}

In this paper, we focus on the computation of pairwise interactions with a cutoff in a grid on a GPU, when there are few particles per cell.
Depending on the interaction kernel, the problem can be memory bound or compute bound.
In the case of a memory-bound problem, the main challenge is to minimize the number of global memory accesses.
Consequently, we aim to develop an efficient implementation using shared memory for this case.

We focus on the interaction kernel, considering that
the input data are the positions of the particles, the number of particles per cell, the starting index of each cell in the sorted array, and the physical values.
The output data are physical values, such as the forces and the potential energy of each particle.
The values are organized in a SoA format, where each component is stored in a separate array.

The width of the cells of the grid is at least equal to the cutoff length, which ensure that the interaction list of a particle must be in the cell where the particle is located or its neighbors.
In the target scenario, the number of particles per cell is bounded, and the distribution of particles is almost uniform.
In case a large portion of the simulation box is empty, or if the particles are clusterized in different parts of the simulation box, most of the approaches would not be adapted because most of them are parallelized over the cells, which would ends in testing all the empty cells to see if there is any work to do.
Consequently, if there are empty cells, they should be removed by working only on the filled part of the simulation box.

%%%%%%%%%%%%%%%%%%%%%%%%%%%%%%%%%%%%%%%%%%%%%%%%%%%%%%%%%%%%%%%%%%%%%%%%%%%%
%%%%%%%%%%%%%%%%%%%%%%%%%%%%%%%%%%%%%%%%%%%%%%%%%%%%%%%%%%%%%%%%%%%%%%%%%%%%
\section{Straightforward Solutions}
\label{sec:classical-solutions}

In this section, we describe various simple and intuitive approaches for which we could not find foundational references. 
Related work is discussed in Section~\ref{sec:related-work}.

\subsection{Naive Implementation (Par-Part-NoLoop)}

The naive implementation consists in using one thread per particle (i.e., parallelization is done over the particles), 
and each thread computes the interactions between its particle (the target) with all the other particles (the sources) of the same cell and its neighbors.
There is no loop over the particles, and no use of the shared memory.
Potentially, threads of the same warp can access the same source particles, which will factorize the memory access, and potentially benefit from the L1 cache.
The global memory accesses are not coalesced when accessing the source particles, but are coalesced when loading the target particle.
The pseudocode of this implementation is provided in Algorithm~\ref{alg:par-part-noloop}.

\begin{algorithm}[h]
    \SetAlgoLined
    \KwIn{Particle data, cell data}
    \KwOut{The $v$ array of the particles is modified}
    \BlankLine
    \textbf{Function} Par-Part-NoLoop(particles, cells)\;
    \Begin{
        idx $\leftarrow$ threadIdx.x + blockIdx.x * blockDim.x\;
        \If{idx $<$ number of particles}{
            part\_target $\leftarrow$ particles[idx]\;
            cell\_idx $\leftarrow$ getCellFromPos(part\_target)\;
            \For{cell\_src $\leftarrow$ cells[getNeighbourList(cell\_idx)]}{
                \ForAll{particles part\_source in cell\_src}{
                    \If{part\_source $\neq$ part\_target}{
                        p2p\_interaction(part\_target, part\_source)\;
                    }
                }
            }
        }
    }
    \caption{Simplified Par-Part-NoLoop implementation.
             The total number of threads instantiated must be equal or greater than the number of particles.}
    \label{alg:par-part-noloop}
\end{algorithm}

\subsection{Naive Flexible Implementation (Par-Part-Loop)}

This implementation is similar to Par-Part-NoLoop, but we use a loop over the target particles.
Consequently, the only difference is that the number of threads or blocks can be arbitrarily selected and not necessarily equal to the number of particles.

\subsection{Cell-based Implementation (Par-Cell)}

In this implementation, we use one thread-block per cell, 
and each thread of this thread-block computes the interaction for a given particle of the cell.
If there are more threads than particles in the cell, some of them will be idle.
There are loops over the cells and the particles, such that the number of threads or blocks can be arbitrary and not necessarily equal to the number of cells or maximum particles per cell.
There is no use of the shared memory, and the global memory accesses are not necessarily coalesced, 
but the source particles are actually the same for all the thread in a thread-block, leading to the same memory accesses.
We provide the pseudocode of this implementation in Algorithm~\ref{alg:par-cell}.

\begin{algorithm}[H]
    \SetAlgoLined
    \KwIn{Particle data, cell data}
    \KwOut{The $v$ array of the particles is modified}
    \BlankLine
    \textbf{Function} Par-Cell(particles, cells)\;
    \Begin{
        \For{cell\_target\_idx $\leftarrow$ blockIdx.x \KwTo number of cells \textbf{step} gridDim.x}{
            cell\_target $\leftarrow$ cells[cell\_target\_idx]\;
            \For{part\_target\_idx $\leftarrow$ threadIdx.x \KwTo cell\_target.size \textbf{step} blockDim.x}{
                part\_target $\leftarrow$ particles[part\_target\_idx + cell\_target.offset]\;
                \For{cell\_src $\leftarrow$ cells[getNeighbourList(cell\_target\_idx)]}{
                    \ForAll{particles part\_source in cell\_src}{
                        \If{part\_source $\neq$ part\_target}{
                            p2p\_interaction(part\_target, part\_source)\;
                        }
                    }
                }
            }
        }
    }
    \caption{Simplified Par-Cell implementation.}
    \label{alg:par-cell}
\end{algorithm}

\subsection{Cell-based Implementation with Shared Memory (Par-Cell-SM)}

This implementation is similar to Par-Cell, but we use shared memory to load the particles of the cells and the neighboring cells.
This allows to access the global memory once for each particle (per thread-block), and then multiple accesses are performed in the shared memory.
The cells are loaded one after the other, and potentially in several steps, such that even if there are many particles per cell,
this approach remains efficient and will not necessarily require a large amount of shared memory 
(ideally the number of particles that can be stored in the shared memory buffer should be a multiple of the number of threads in the thread-block).
Algorithm~\ref{alg:par-cell-sm} provides the pseudocode of his implementation.

\begin{algorithm}[H]
    \SetAlgoLined
    \KwIn{Particle data, cell data, size of the shared memory}
    \KwOut{The $v$ array of the particles is modified}
    \BlankLine
    \textbf{Function} Par-Cell(particles, cells, sm\_size)\;
    \Begin{
        \_\_shared\_\_ sm\_particles[sm\_size]\;
        \For{cell\_target\_idx $\leftarrow$ blockIdx.x \KwTo number of cells \textbf{step} gridDim.x}{
            cell\_target $\leftarrow$ cells[cell\_target\_idx]\;
            \For{part\_target\_idx $\leftarrow$ threadIdx.x \KwTo cell\_target.size + blockDim.x - 1 \textbf{step} blockDim.x}{
                thread\_compute $\leftarrow$ (part\_target\_idx $<$ cell\_target.size)\;
                \If{thread\_compute}{
                    part\_target $\leftarrow$ particles[part\_target\_idx + cell\_target.offset]\;
                }
                \For{cell\_src $\leftarrow$ cells[getNeighbourList(cell\_target\_idx)]}{
                    \For{part\_src\_idx $\leftarrow$ 0 \KwTo cell\_src.size \textbf{step} blockDim.x}{
                        copy\_particles(sm\_particles, particles, cell\_src, part\_src\_idx)\;
                        \If{thread\_compute}{
                            \ForAll{particles part\_source in sm\_particles}{
                                \If{part\_source $\neq$ part\_target}{
                                    p2p\_interaction(part\_target, part\_source)\;
                                }
                            }
                        }
                    }
                }
            }
        }
    }
    \caption{Simplified Par-Cell-Sm implementation.}
    \label{alg:par-cell-sm}
\end{algorithm}

%%%%%%%%%%%%%%%%%%%%%%%%%%%%%%%%%%%%%%%%%%%%%%%%%%%%%%%%%%%%%%%%%%%%%%%%%%%%
%%%%%%%%%%%%%%%%%%%%%%%%%%%%%%%%%%%%%%%%%%%%%%%%%%%%%%%%%%%%%%%%%%%%%%%%%%%%
\section{Proposed Solutions}
\label{sec:proposed-solutions}

We propose two approaches to perform the computation while using shared-memory in a more advanced way: full load (All-in-SM) and pencil load (X-pencil).
In both cases, the idea is to copy particles in shared memory and use them to compute several interactions, 
instead of loading one particle for each interaction.
A main difference with the Par-Cell-SM approach is that we consider that there are few particles per cell, hence we aim at loading several cells in shared memory together, i.e. a thread-block should manage more than one cell.

%%%%%%%%%%%%%%%%%%%%%%%%%%%%%%%%%%%%%%%%%%%%%%%%%%%%%%%%%%%%%%%%%%%%%%%%%%%%
\subsection{Full Load (All-in-SM)}

In this strategy, we aim to maximize the use of the particles loaded in shared memory.
The pattern is as follows.
We know that, aside from the border cells, each cell is accessed by $3^3$ cells (including self-cell interaction), as are the particles.
This also applies to a sub-box of cells, where each internal cell is used by $3^3$ cells.
The cells on the border of the sub-box are used between $9$ and $1$ times, as they are ghost cells (they only contain source particles).
Therefore, we can load a sub-box into shared memory to leverage the repeated reuse of the cells.
The size of the sub-box that can be loaded is limited by the amount of shared memory available per thread-block.
Since the grid/thread configuration and the size of shared memory are fixed at kernel launch, we need to determine the dimension of the sub-box before invoking the kernel.
To achieve this, we retain the maximum number of particles in a cell when computing the prefix sum of the number of particles per cell, denoted as $M_C$.
Given that we know the number of floating-point values required per particle, we deduce the maximum memory occupancy of a cell in shared memory, denoted as $M_{C*}$, and request this size for each cell we intend to load as dynamic shared memory at kernel invocation time.
This also implies that a portion of the shared-memory buffer might remain unused if one or more cells contain fewer than $M_C$ particles. Consequently, we expect that the effectiveness of the method will be tied to the uniformity of the distribution and the difference between the maximum ($M_C$) and the average number of particles per cell.

\begin{figure}
    \centering

    \begin{subfigure}[b]{0.38\textwidth}
        \includegraphics[width=\textwidth, height=0.23\textheight, keepaspectratio]{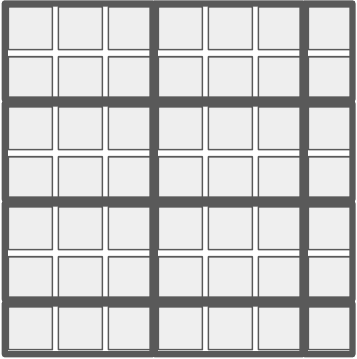}
        \caption{2D example of sub-box configuration.}
        \label{subfig:cell-config}
    \end{subfigure}
    \hfill
    \begin{subfigure}[b]{0.58\textwidth}
        \centering
        \includegraphics[width=\textwidth, height=0.3\textheight, keepaspectratio]{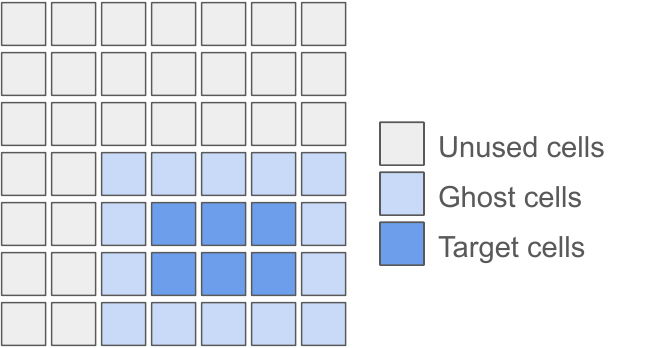}
        \caption{Example of the cells used by a sub-box.}
        \label{subfig:sub-box-config}
    \end{subfigure}
    
    \caption{2D example of sub-box configuration.
            In this case, consider that the maximum number of cells that can be loaded is $5\times4=20$.
            As it includes the ghost cells, the final target cells form a rectangle of $3\times2$ cells (Figure~\ref{subfig:cell-config}).
            Each rectangle (sub-box in 3D) is assigned to a thread-block and is loaded with its neighbors in shared memory (Figure~\ref{subfig:sub-box-config}).
            We have the guarantee that any cell contains at most $M_C$ particles, so we allocated enough memory to store $20 \times M_C$ particles.}
    \label{fig:sub-box-config}
\end{figure}

To find the dimension of the sub-box we can assign to a thread-block, we first divide the size of the available shared memory by $M_{C*}$, which gives us the maximum cells we can load.
This number is not necessarily a power of 3.
Therefore, we find the closest power of 3 that is not greater than the memory limit, called $p_3$.
The final size of the sub-box $B_{X \times Y \times Z}$ is the greatest value among $p_3^3$, $p_3+1 \times p_3^2$, $p_3+1 \times p_3+1 \times p_3$ or $p_3+2 \times p_3^2$, 
which is not greater than the memory limit.
This size includes the ghost cells and leads to a sub-box that is not necessarily a cube.
As a remark, if $B_{X \times Y \times Z} < 27$, this approach cannot be used, as it is impossible to load one target cell and its ghost neighbors together.
A possible configuration is provided in Figure~\ref{fig:sub-box-config}.

We create as many thread-blocks as there are sub-boxes in the simulation domain, 
avoiding iterating over the sub-boxes, while potentially increasing the occupancy.
The number of threads per thread-block is usually $M_C \times B_{X \times Y \times Z}$, which is the maximum number of particles in the sub-box.
Therefore, each thread copies only one particle from the global memory to the shared memory buffer.
The memory accesses are not necessarily fully coalesced because the position of the first particle is not aligned with the size of a warp.
The pseudocode of this implementation is provided in Algorithm~\ref{alg:all-in-sm}.

\begin{algorithm}[H]
    \SetAlgoLined
    \KwIn{Particle data, cell data, size of the shared memory}
    \KwOut{The $v$ array of the particles is modified}
    \BlankLine
    \textbf{Function} All-in-SM(particles, cells, sm\_size)\;
    \Begin{
        \_\_shared\_\_ sm\_particles[sm\_size]\;
        sub\_box\_config $\leftarrow$ computeSubBoxConfig(gridDim, blockIdx)\;
        copy\_particles(sm\_particles, particles, sub\_box\_config)\;

        cell\_target $\leftarrow$ cells[threadIdx.x/$M\_C$]\;
        \For{part\_target\_idx $\leftarrow$ threadIdx.x \KwTo cell\_target.size \textbf{step} blockDim.x}{
            part\_target $\leftarrow$ sm\_particles[part\_target\_idx + cell\_target.offset]\;
            \For{cell\_src $\leftarrow$ cells[getNeighbourList(cell\_target)]}{
                \ForAll{particles part\_source in sm\_particles[cell\_src]}{
                    \If{part\_source $\neq$ part\_target}{
                        p2p\_interaction(part\_target, part\_source)\;
                    }
                }
            }
        }
    }
    \caption{Simplified All-in-SM implementation.}
    \label{alg:all-in-sm}
\end{algorithm}

It is important to understand that this algorithm actually needs several arrays, all stored in shared memory.
The first one is the array of global offset (the prefix sum array), which is simply a copy of the corresponding array from global memory.
Because these numbers are used several times in the computation, it is better to have them in shared memory.
The second one is an array of local offset, which allows knowing the offset of the cells in the shared memory buffer.
The third one is a one-to-one mapping between the particles' indices in the shared memory arrays and their corresponding indices in the global memory.
The fourth one is the array of the particles' values in the shared memory.

\begin{figure}
    \centering
    \includegraphics[width=\textwidth, height=0.3\textheight, keepaspectratio]{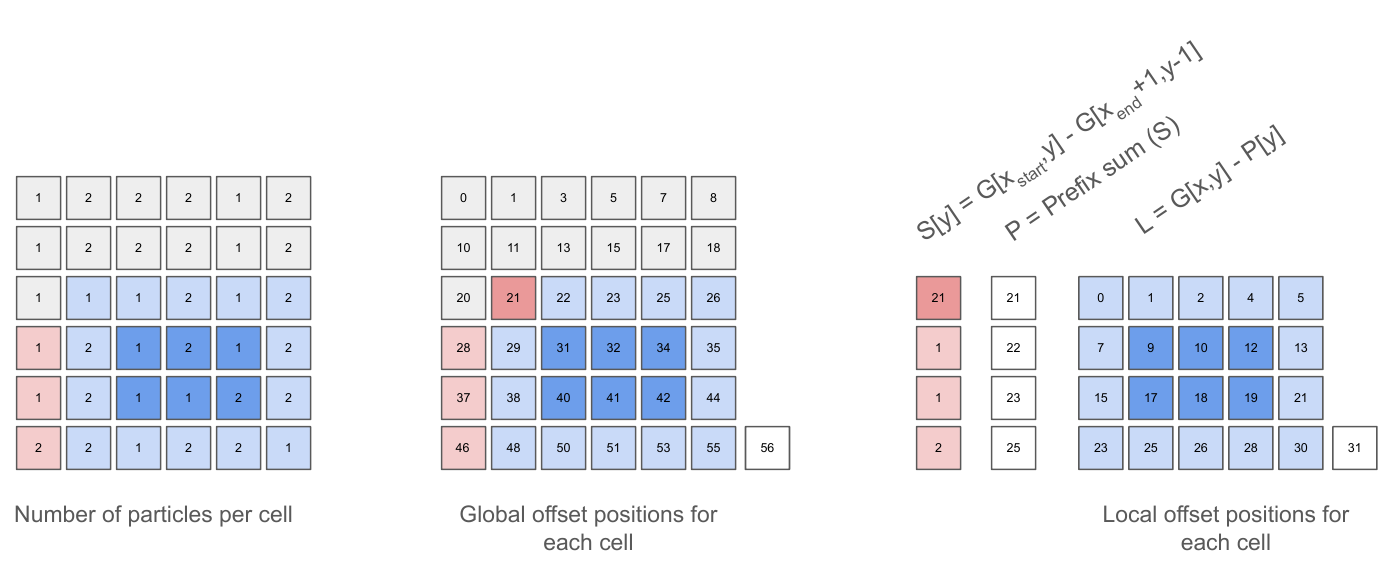}
    \caption{2D example of a local offset computation.}
    \label{fig:local-prefix}
\end{figure}

Building the local offset array is done in two steps, as shown in Figure~\ref{fig:local-prefix}.
First, we need to compute the number of particles inside each X-pencil of the sub-block (an X-pencil is a sub-box with its size greater than one only in the X dimension).
This can be done by computing the difference between the global offset of the first cell of the X-pencil, and the offset of the cell after the X-pencil.
This is valid even if this last cell is actually at the next Y or Z coordinate due to linearization.
Once we have this array, we put as first value the offset of the corner cell.
Then, we perform a prefix sum on this array such that we obtain the values to be removed from the global offset to obtain the local offset 
(each value represents the particles that are not included in the sub-box but which are counted in the global offset).
Efficient prefix sum algorithm in shared memory is described in Section~\ref{sec:prefixum}.

While this approach aims at maximizing the shared memory reuse, the amount of shared memory it uses is high and a clear limitation.
For instance, if there are a maximum of 16 particles per cell, and we need 8 single precision floating point values per particle, 
we need 512 bytes of shared memory per cell.
If we have 48KB of shared memory, we can load 96 cells, which is a 4x4x6 sub-box (and active 2x2x4 = 16 cells without the ghosts).
And doing so we would have only a single active block per multiprocessor because one thread-block will use all the shared memory.
Even if we have a single active cell, we need 27 cells in shared memory, which is a 3x3x3 sub-box and will take 13.5KB, and will allow having only 3 active blocks per multiprocessor.
Consequently, the shared memory usage is expected to have a strong impact on the occupancy of the GPU.

The memory accesses are potentially not coalesced because the position of the first particle of the X-pencil is not aligned with the thread index,
as the number of particles in the previous cells is not necessarily a multiple of the number of threads in a warp.

%%%%%%%%%%%%%%%%%%%%%%%%%%%%%%%%%%%%%%%%%%%%%%%%%%%%%%%%%%%%%%%%%%%%%%%%%%%%
\subsection{Pencil Load (X-pencil)}

\begin{figure}
    \centering
    \includegraphics[width=0.7\textwidth, height=\textheight, keepaspectratio]{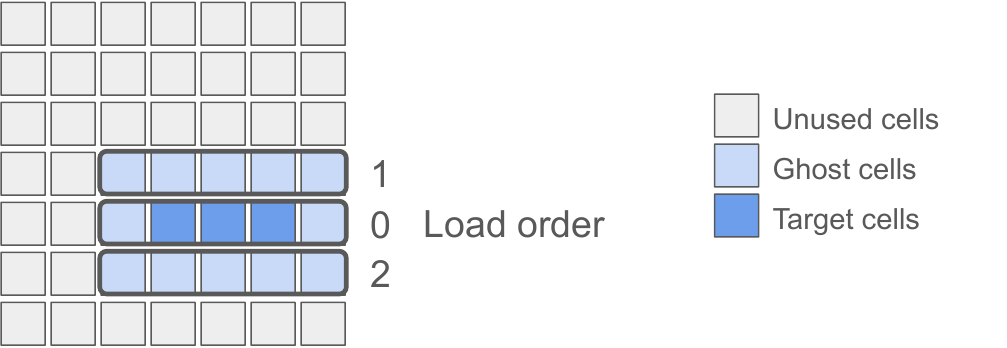}
    \caption{2D example of the X-pencil.
            First, the X-pencil that covers the target particles is loaded in shared memory, and use for computation.
            Then, a loop in Y/Z will load the other pencils, one at a time, and use them for computation.}
    \label{fig:example-x-pencil}
\end{figure}

This approach can be seen as a variant of the All-in-SM, where the sub-block is a pencil.
In addition, we do not load all the particles/cells that will be used by the computation together, but we load them in several steps, as shown in Figure~\ref{fig:example-x-pencil}.
Again, the size of the pencil is limited by the amount of shared memory available per thread-block,
consequently the length of the X-pencil is $SM / (M_C \times sizeof(Part))$, considering that $sizeof(Part)$ is the memory size of a particle.
This length, will include the ghost cells at the extremities, so if it is lower than 3, 
it would mean that this method cannot be applied because an X-pencil of 3 cells cannot be loaded.

In this approach, there must be at least one thread per particle, because the target particles/cells are not kept in the shared memory but only in the registers of the threads.
Therefore, this method would not work if they were more particles in the X-pencil that the number of threads we can have in a thread-block.

We provide an overview of our approach in Algorithm~\ref{alg:x-pencil}.
First, we load the particles of the X-pencil into shared-memory, and each thread puts its particle into its registers.
Then, each thread computes the interactions with the cell that includes its particle and the neighbor cells that are on the same X-pencil.
Finally, we iterate along axes Y and Z (by -1/+1), and load first the particles of the ghost cells, and then compute the interactions.

The memory accesses are potentially not coalesced because the position of the first particle of the X-pencil is not aligned with the size of warp.
This approach requires less shared memory than All-in-SM and does not need to build complex arrays using a prefix sum in shared memory, 
which simplifies its implementation.

\begin{algorithm}
    \SetAlgoLined
    \KwIn{Particle data, cell data, size of the shared memory}
    \KwOut{The $v$ array of the particles is modified}
    \BlankLine
    \textbf{Function} All-in-SM(particles, cells, sm\_size)\;
    \Begin{
        \_\_shared\_\_ sm\_particles[sm\_size]\;
        sub\_box\_config $\leftarrow$ computeSubBoxConfig(gridDim, blockIdx)\;
        copy\_particles(sm\_particles, particles, sub\_box\_config, blockIdx.y, blockIdx.z)\;

        nb\_particles\_to\_compute $\leftarrow$ get\_nb\_particles(sub\_box\_config.x\_interval, cells)\;
        has\_to\_compute $\leftarrow$ (threadIdx.x $<$ nb\_particles\_to\_compute)\;
        \_\_syncthreads()\;
        \If{has\_to\_compute}{
            my\_particle $\leftarrow$ sm\_particles[sub\_box\_config.x\_pencil\_offset + threadIdx.x]\;
            compute\_interactions(my\_particle, sm\_particles, cells, has\_to\_compute)\;
        }
        \_\_syncthreads()\;

        \For{Z $\leftarrow$ Max(0, blockIdx.z-1) \KwTo Min(sub\_box\_config.z\_limit, blockIdx.z+2)}{
            \For{Y $\leftarrow$ Max(0, blockIdx.y-1) \KwTo Min(sub\_box\_config.y\_limit, blockIdx.y+2)}{
                \If{Y $\neq$ blockIdx.y \textbf{or} Z $\neq$ blockIdx.z}{
                    copy\_particles(sm\_particles, particles, sub\_box\_config, Y, Z)\;
                    \_\_syncthreads()\;

                    \If{has\_to\_compute}{
                        compute\_interactions(my\_particle, sm\_particles, cells, has\_to\_compute)\;
                    }
                    \_\_syncthreads()\;
                }
            }
        }
    }
    \caption{Simplified X-pencil implementation.}
    \label{alg:x-pencil}
\end{algorithm}

The memory accesses are potentially not coalesced because the position of the first particle of the X-pencil is not aligned with the size of warp.
This approach requires less shared memory than All-in-SM and does not need to build complex arrays using a prefix sum in shared memory, 
which simplified its implementation.

%%%%%%%%%%%%%%%%%%%%%%%%%%%%%%%%%%%%%%%%%%%%%%%%%%%%%%%%%%%%%%%%%%%%%%%%%%%%
\subsection{Pencil load with register storage (X-pencil-reg)}

We added a variant of the X-pencil approach, where the particles are first stored in the registers of the threads. 
This approach is not faster than the X-pencil approach in practice, but it is tied to an interesting programming pattern 
on GPU that we aim to present (the performance results are not presented in the current manuscript).

In the X-pencil, the X-pencil that cover the target cells is loaded in shared memory, and each thread puts its particle into its registers. 
Then, the computation is performed using the previously loaded X-pencil.
After that, we iterate along axes Y and Z (by -1/+1), and load first the particles of the ghost cells, and then compute the interactions.
So for any X-pencil that is loaded, all the target cells/particles are involved in the computation.
The X-pencil strategy is limited by the maximum number of particles that are in an X-pencil as all of them should fit in shared memory. 
For $L$ target cells, we need to load $3^2$ pencils of length $L + 2$ cells (the plus two is for the ghost cells). 

In the X-pencil-reg, the target particles are first loaded in registers. 
Then, we load all the needed X-pencil, one after the other, in shared memory and perform the interactions. 
However, we avoid accessing the global memory for all the particles that will be used as sources but which are included 
in the target sub-box because they are already in registers.
Consequently, it can be copied to the shared memory at the right iteration. 
Said differently, the registers are used as temporary  storage for the target particles,
and the shared memory is used to store the source particles (which can also be targets).

\begin{figure}
    \centering
    \includegraphics[width=0.9\textwidth, height=\textheight, keepaspectratio]{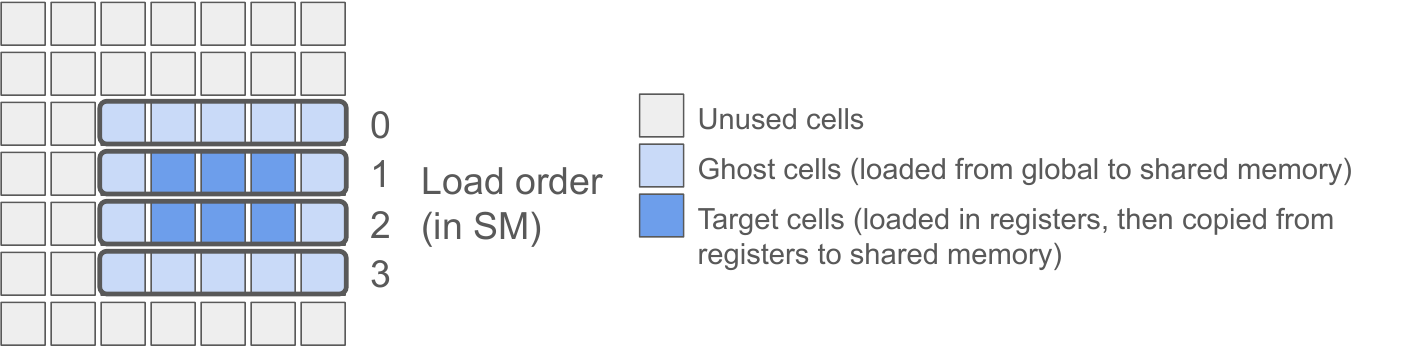}
    \caption{2D example of the X-pencil-reg.
            First, the target particles are loaded in registers.
            Then, the source particles are loaded in shared memory (one pencil after the other).
            When possible, the source particles are copied from the registers.
            After each load of a pencil, the interactions are computed.
            It is possible that some target cells are not involved in the computation for some iterations, and the threads are idle, depending on the pencil's position.}
    \label{fig:example-x-pencil-reg}
\end{figure}

The main advantage is that instead of having target cells organized as a pencil in the X direction,
we can target a sub-box of cells.
This allows to have a better occupancy and/or a better re-use of the data loaded from the global memory in shared memory or registers.
An overview of the method is provided in Figure~\ref{fig:example-x-pencil-reg}.

%%%%%%%%%%%%%%%%%%%%%%%%%%%%%%%%%%%%%%%%%%%%%%%%%%%%%%%%%%%%%%%%%%%%%%%%%%%%
%%%%%%%%%%%%%%%%%%%%%%%%%%%%%%%%%%%%%%%%%%%%%%%%%%%%%%%%%%%%%%%%%%%%%%%%%%%%
\section{Prefix-sum GPU Implementation on Shared Memory}
\label{sec:prefixum}

In our full load strategy (All-in-SM), we need to compute the prefix sum of the number of particles per cell.
Most implementations are based on the Blelloch's algorithm, 
which is a simple and efficient algorithm to compute the prefix sum on a vector processing unit.
Both Blelloch's algorithm and ours are based on the idea of building a binary tree of sums, where each node is the sum of its two children, and consists of an upward pass followed by a downward one.
In the upward pass, we compute the prefix sum of the array by building the binary tree of sums.
In the downward pass, we compute the prefix sum of the array by traversing the binary tree of sums from the root to the leaves.
The complexity of the sequential algorithm is $O(N)$, where $N$ is the size of the array.
With $N/2$ threads, each will execute an algorithm with a complexity of $O(\log N)$.

Most implementations are designed as GPU kernel that should be called from the host~\cite{nguyen2007gpu}.
The data to be processed are located in the global memory, and the results are stored in the global memory.
In our case, we need to compute the prefix sum on shared memory, and we need to store the results in shared memory.
Consequently, we propose an implementation of the prefix sum on shared memory, 
that minimizes the number of memory accesses and do not need extra shared memory.
Our implementation build a prefix tree but put the numbers at the right places during the upward pass, 
and then during the downward pass, we compute the prefix sum of the other elements.

We provide our approach in Algorithm~\ref{alg:prefix-sum}. 
In the upward pass, we simply accumulate values as if we were building the upper levels of the tree, even if we do it in place.
First, we sum the values at odd indices into their left neighbors at even indices.
Then, we repeat this process with the elements at odd/even indices from the previous iteration. 
This is why the step variable, called \texttt{js} in the algorithm, is multiplied by two at each iteration.
We stop when there are no more elements to sum. 
For an array of 8 elements \texttt{1.1.1.1.1.1.1.1.1.1}, the results after the inner loop would be \texttt{\textbf{1}.\textbf{\underline 2}.\textbf{1}.\textbf{\underline 2}.\textbf{1}.\textbf{\underline 2}.\textbf{1}.\textbf{\underline 2}.\textbf{1}.\textbf{\underline 2}}, \texttt{1.\textbf{2}.1.\textbf{\underline 4}.1.\textbf{2}.1.\textbf{\underline 4}} and \texttt{1.2.1.\textbf{4}.1.2.1.\textbf{\underline 8}}.

In the downward pass, we propagate the results previously computed by the upward pass.
Each node's computed sum is added to its right child, except for the last node of each level.
Consequently, at each outer iteration, we divide the step and shift variables by two, and we continue until the shift step becomes $1$.
For our example, this would give \texttt{1.2.1.\textbf{4}.1.\textbf{\underline 6}.1.8} and \texttt{1.\textbf{2}.\textbf{\underline 3}.\textbf{4}.\textbf{\underline 5},\textbf{6}.\textbf{\underline 7}.8}.

If we consider an array of $N$ elements, the abstract binary tree that can cover it has a height of $h = \lceil \log_2(N + 1) \rceil$.
Here, $h$ is the number of levels, such that if the root is at level $0$, the leaves are at level $h-1$. 
Our upward pass starts from $h-2$ up to level $0$, and our downward pass starts at level $2$ down to level $h-1$. 
Consequently, there are a total of $2 \times h - 3$ synchronizations, instead of $2 \times h$ in Blelloch's algorithm.
The total number of nodes updated during the upward pass is $N-1$, and $N-h$ during the downward pass.
Each update consists of one read and one read-write.
In our example, there are 5 synchronizations and 10 nodes updated.

Finally, in the case the inner loops are not parallelized by the threads with a loop, but instead consist of a simple test to check if the element assigned to a thread is less than $N$ (i.e., we need at least as many threads as the number of element to proceed in the largest inner loops), our approach can require half the number of threads compared to Blelloch's algorithm.

\begin{algorithm}[htb]
    \SetAlgoLined
    \KwIn{prefix array of type NumType, N of type IndexType}
    \KwOut{The prefixTree array is modified to contain prefix sums}
    \BlankLine
    \textbf{Function} buildPrefixSMDevice(prefix, N)\;
    \Begin{       
        \tcc{Upward pass}
        js $\leftarrow$ 2\;
        \While{js $\leq$ N}{
            jsd2 $\leftarrow$ js / 2\;
            \For{idN $\leftarrow$ js - 1 \KwTo N-1 \textbf{step} js}{
                prefix[idN] $\leftarrow$ prefix[idN] + prefix[idN - jsd2]\;
            }
            js $\leftarrow$ js * 2\;
        }
        
        \tcc{Downward pass}
        js $\leftarrow$ max(4, js / 4)\;
        \While{js $>$ 1}{
            jsd2 $\leftarrow$ js / 2\;
            \For{idN $\leftarrow$ js + jsd2 - 1 \KwTo N-1 \textbf{step} js}{
                prefix[idN] $\leftarrow$ prefix[idN] + prefix[idN - jsd2]\;
            }
            js $\leftarrow$ jsd2\;
        }
    }
    \caption{Sequential Prefix Sum Calculation. A parallel implementation is obtained by parallelizing the inner loops and adding a synchronization just after.}
    \label{alg:prefix-sum}
\end{algorithm}

It seems that this algorithm has already been proposed in the literature, but we could not find a reference.

%%%%%%%%%%%%%%%%%%%%%%%%%%%%%%%%%%%%%%%%%%%%%%%%%%%%%%%%%%%%%%%%%%%%%%%%%%%%
%%%%%%%%%%%%%%%%%%%%%%%%%%%%%%%%%%%%%%%%%%%%%%%%%%%%%%%%%%%%%%%%%%%%%%%%%%%%
\section{Experiments}
\label{sec:experiments}

\subsection{Experimental Setup}

We performed the experiments on three different platforms:
\begin{itemize}
    \item \emph{NVIDIA T600} with 4GB GDDR6, 48KB of shared-memory, 40 multiprocessors, and using NVCC V12.0.76;
    \item \emph{NVIDIA A100} with 40GB hBM2, 48KB of shared-memory, 108 multiprocessors, and using NVCC V12.3.107.
    \item \emph{AMD Instinct MI210} with 64GB hBM2, 64KB of shared-memory, 104 multiprocessors, and using HIPCC V5.7.31921 (ROC 5.7.2).
          The HIP code has been obtained by a simple translation of the CUDA code using hipify-perl.~\footnote{https://rocm.docs.amd.com/projects/HIPIFY/en/latest/hipify-perl.html}
\end{itemize}

The code was compiled with the following flags: \texttt{-O3 -arch=sm\_75} for the T600, \texttt{-O3 -arch=sm\_80} for the A100, and \texttt{-O3 --offload-arch=gfx90a} for the MI210 (and \texttt{-DNDEBUG} on all).
For each kernel, we execute 200 asynchronous calls and take the execution time of all of them divided by 200 as reference.
We use a simulation grid with $3^d$ number of cells, with $d$ being $2$, $4$, $8$, $16$ or $32$, and we put 1, 10 or 100 particles per cell in average.
The particles are uniformly distributed in the simulation domain, such that empty cells are very unlikely.
We perform the computation in single floating point precision.

The CUDA grid/thread-block configurations are the following:
\begin{itemize}
    \item Par-Part-NoLoop: We use 128 threads per thread-block, and as many blocks as needed to have one thread per particle;
    \item Par-Part-Loop: We use the same configuration as Par-Part-NoLoop, so we expect to see no difference;
    \item Par-Cell: We use 128 threads per thread-block, and as many blocks as there are cells;
    \item Par-Cell-SM: We use the same configuration as Par-Cell, and a shared memory of 512 particles per thread-block;
    \item All-in-SM: We use $M_C \times B_{X \times Y \times Z}$ threads, which is the maximum number of particles in the sub-box.
                     However, we limit this number to 512.
                     The number of blocks is the number of sub-boxes in the simulation domain.
                     We remind that the size of a sub-block is limited by the amount of shared memory available per thread-block, 
                    and in case the number of blocks is less than the number of processors, we reduced the size of the sub-block to ensure enough
                    parallelism. 
                    For example, if we have enough shared memory to have $B = {2, 2, 2}$ but then have only 12 blocks for a GPU with 40 processors,
                    we will reduce the sub-block to $B = {2, 1, 1}$ (obtaining 48 blocks);
    \item X-pencil: We use $M_C \times (L_X+2)$ threads, which is the maximum number of particles in the X-pencil, including the ghost cells.
                    We limit this number to 1024 (there must be less than 1024 particles in the X-pencil).
                    The number of blocks is the number of X-pencil in the simulation domain.
                    We remind that the size of a X-pencil is limited by the amount of shared memory available per thread-block, 
                    and in case the number of blocks is less than the number of processors, we reduce the length of the X-pencil to ensure enough
                    parallelism.
                    For example, if for a given configuration, we have enough shared memory to have $L_X = 2$ and have 12 blocks for a GPU with 40 processors,
                    we will reduce the length of the X-pencil to 1 (obtaining 24 blocks).
\end{itemize}

The interaction kernel is the Lennard-Johns potential:
\begin{equation}
    \text{K}(i, j) = 4 \times E_0 \times \bigg( \left( \frac{d(i,j)}{r} \right)^{12} - \left( \frac{d(i,j)}{r} \right)^6 \bigg).
\end{equation}
Since the particles are randomly distributed, we use a softening factor to avoid singularities.
This kernel has an arithmetic intensity of $0.4$ FLOP/byte (if we considered that a square root cost is 1 FLOP, giving a total of 21 FLOP per interaction).

%%%%%%%%%%%%%%%%%%%%%%%%%%%%%%%%%%%%%%%%%%%%%%%%%%%%%%%%%%%%%%%%%%%%%%%%%%%%

\subsection{Results}

\begin{figure}
    \centering
    \begin{subfigure}[b]{0.48\textwidth}
        \centering
        \includegraphics[width=0.4\textwidth]{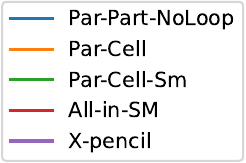}
        \vspace{2cm}
    \end{subfigure}
    \hfill
    \begin{subfigure}[b]{0.48\textwidth}
        \centering
        \includegraphics[width=\textwidth]{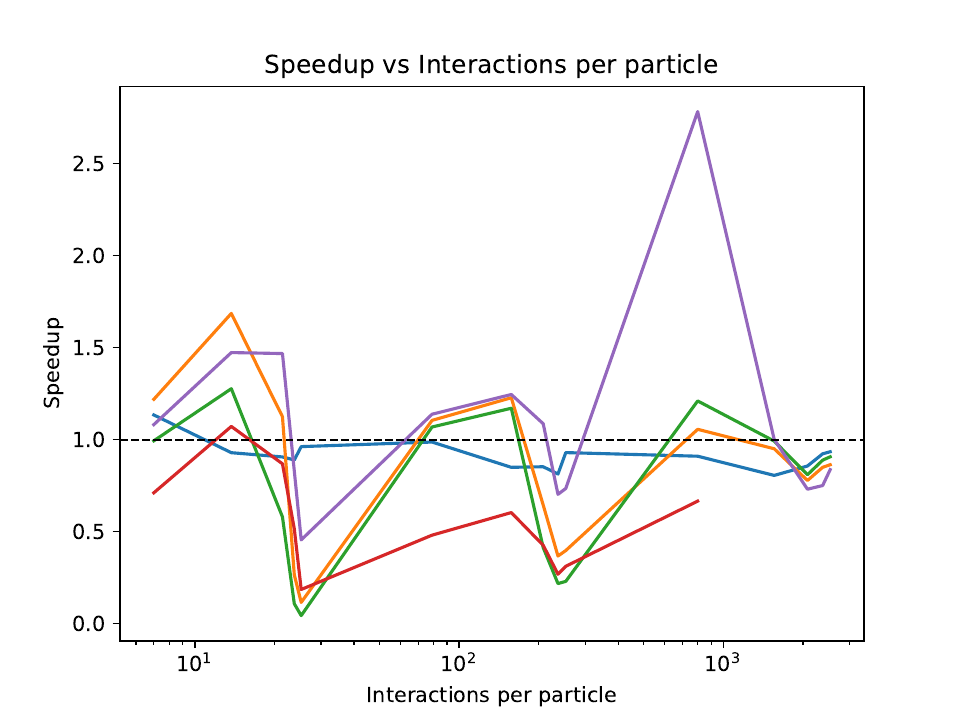}
        \caption{T600.}
        \label{fig:res:laptop}
    \end{subfigure}

    \begin{subfigure}[b]{0.48\textwidth}
        \centering
        \includegraphics[width=\textwidth]{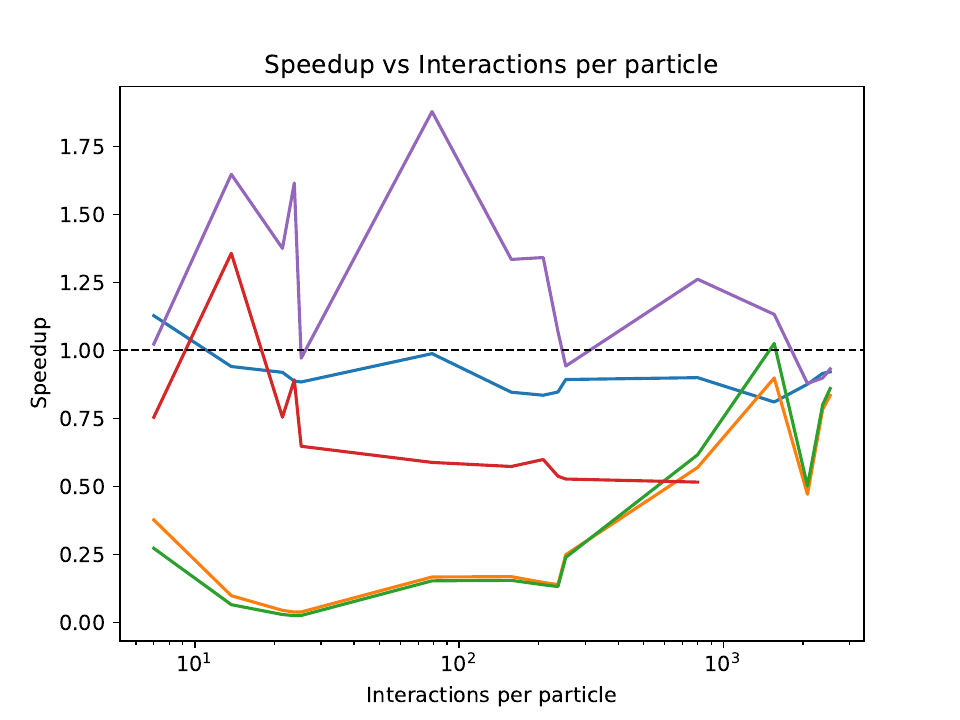}
        \caption{A100.}
        \label{fig:res:a100}
    \end{subfigure}
    \hfill
    \begin{subfigure}[b]{0.48\textwidth}
        \centering
        \includegraphics[width=\textwidth]{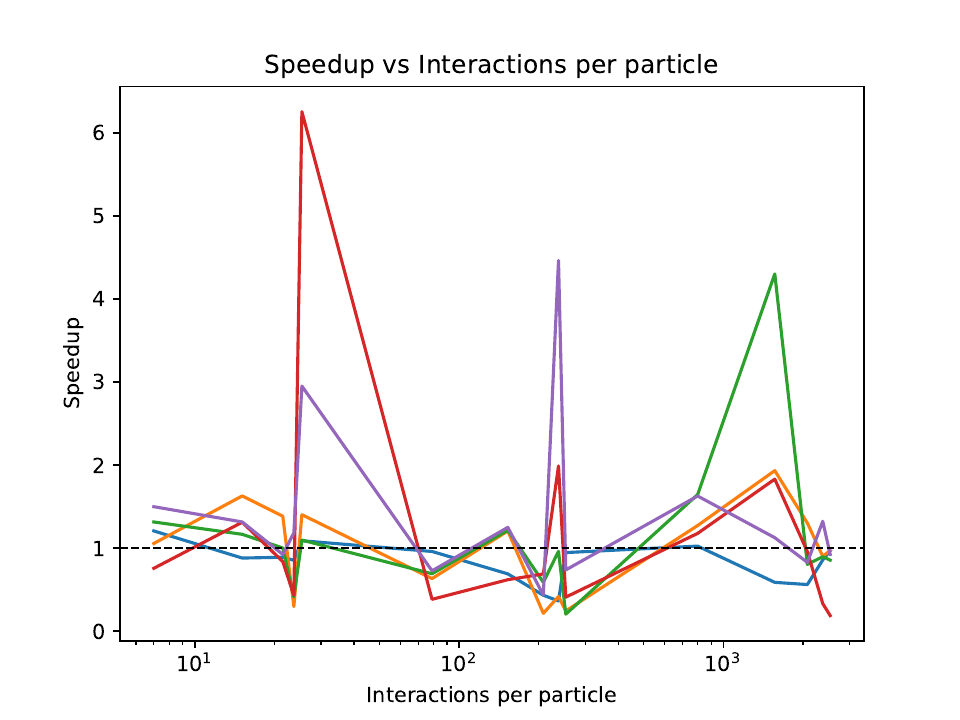}
        \caption{MI210.}
        \label{fig:res:mi210}
    \end{subfigure}

    \caption{Performance results: speedup against Par-Part-NoLoop.
             The $X$ axis corresponds to the total number of interactions computed divided by the number of particles.
             It also matches the number of particles per cell (1, 10, 100) and the simulation box division (from 2 to 32 by power of 2).}
    \label{fig:perf}
\end{figure}

Figure~\ref{fig:perf} provides the results of the experiments, with Figure~\ref{fig:res:laptop} for the T600, 
Figure~\ref{fig:res:a100} for the A100, and Figure~\ref{fig:res:mi210} for the MI210.
On NVidia architectures, the classical solutions (Par-Part-Loop, Par-Cell and Par-Cell-SM) 
are not competitive against Par-Part-NoLoop.
Par-Part-Loop is providing very close performance, as expected because the only difference is that a $if$ is replaced by a $for$.
Concerning Par-Cell, it suffers from the parallelization over the cells, 
because the number of particles per cell is not necessarily a multiple of the thread warp,
which causes bad memory accesses and leads to idle threads in the last warp of each thread-block.
This is highly visible with few interactions (and thus few particles per cell).
The use of shared-memory does not provide a significant benefit, except when there are lots of particles.
This demonstrates that this classical pattern of loading the particles in a shared memory buffer, one cell after the other,
should be used only with a really large number of particles per cell (above 100).

Looking at the All-In-SM, we can see that it is not competitive against Par-Part-NoLoop.
We expect that the overuse of shared memory leads to a poor GPU occupancy.
Moreover, there are too many particles per cells when the number of interactions is above $10^3$, and so that the method becomes unusable.

Concerning X-pencil, it can provide a performance gain in several cases, but also a slight slowdown in few other cases.
On T600, the speedup can be as high as $\times 2.5$, which is significant, and can be $\times 1.7$ on the A100.
However, on A100 the speedup is more frequent. 

On the AMD architecture, the X-pencil is providing a significant speedup on most cases and with a peak of $\times 1.65$.
The Par-Cell-SM becomes competitive as the number of particles increases.
The All-in-SM is not competitive, except for one case.
This shows that even if the MI210 has a larger amount of shared memory compared to the NVidia architectures, 
it is still not enough to make this approach competitive.

\begin{figure}[htb]
    \centering
    \begin{subfigure}[b]{\textwidth}
        \centering
        \includegraphics[width=.2\textwidth, height=0.2\textheight, keepaspectratio]{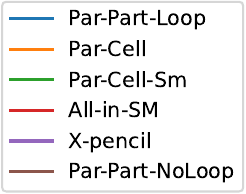}
    \end{subfigure}
    \hfill

    \begin{subfigure}[b]{0.48\textwidth}
        \centering
        \includegraphics[width=\textwidth, height=0.2\textheight, keepaspectratio]{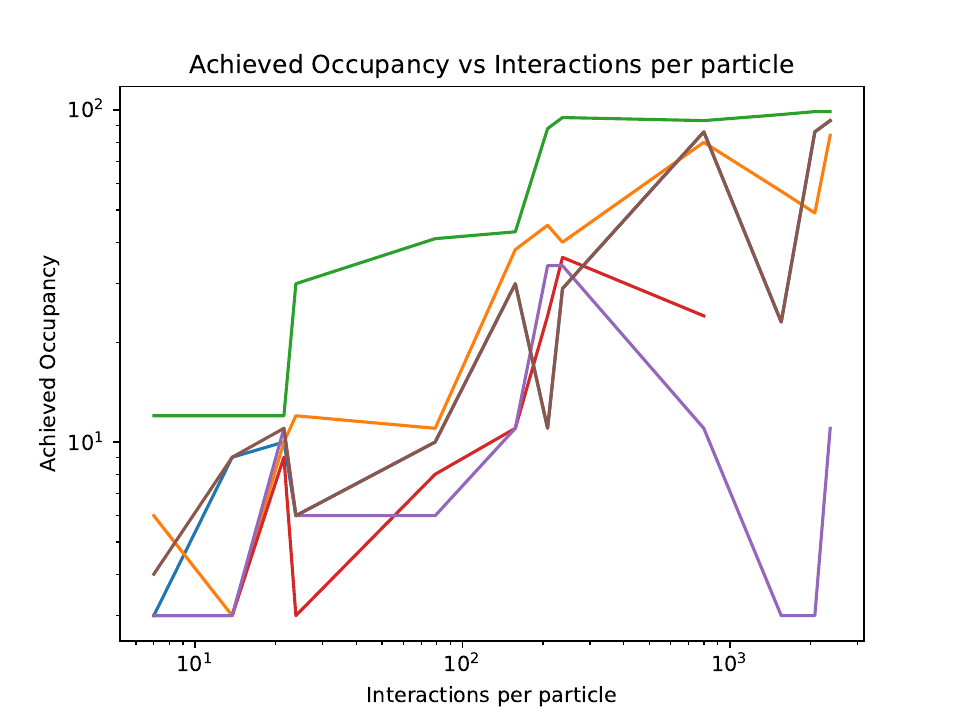}
        \caption{Achieved occupancy (\%).}
        \label{fig:res:achieveoccupancy}
    \end{subfigure}
    \hfill  
    \begin{subfigure}[b]{0.48\textwidth}
        \centering
        \includegraphics[width=\textwidth, height=0.2\textheight, keepaspectratio]{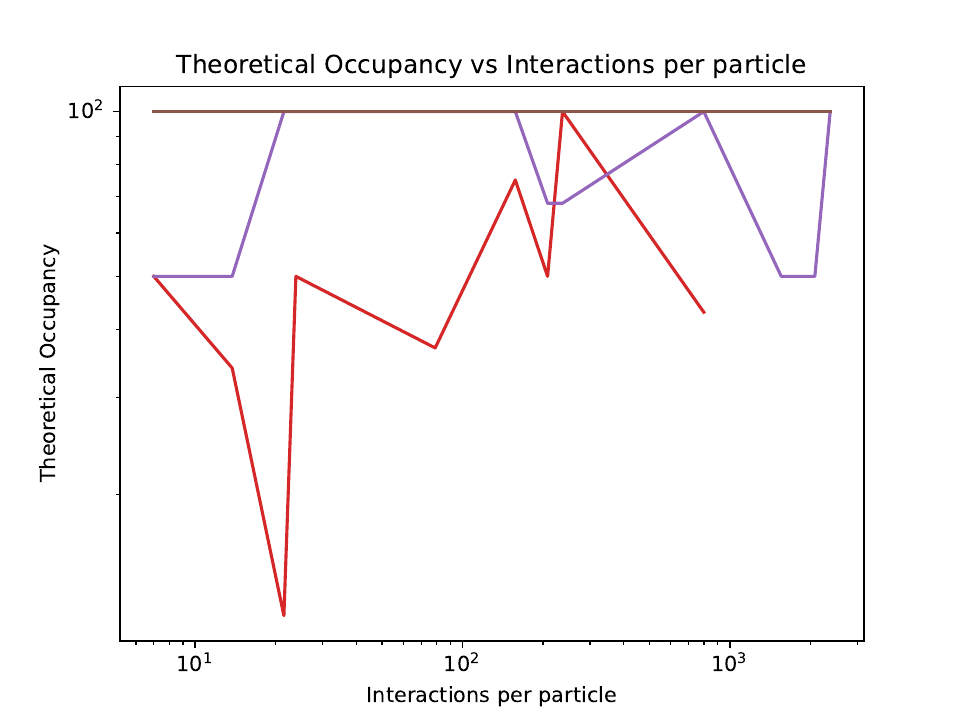}
        \caption{Theoretical occupancy (\%).}
        \label{fig:res:theoreticaloccupancy}
    \end{subfigure}

    \begin{subfigure}[b]{0.48\textwidth}
        \centering
        \includegraphics[width=\textwidth, height=0.2\textheight, keepaspectratio]{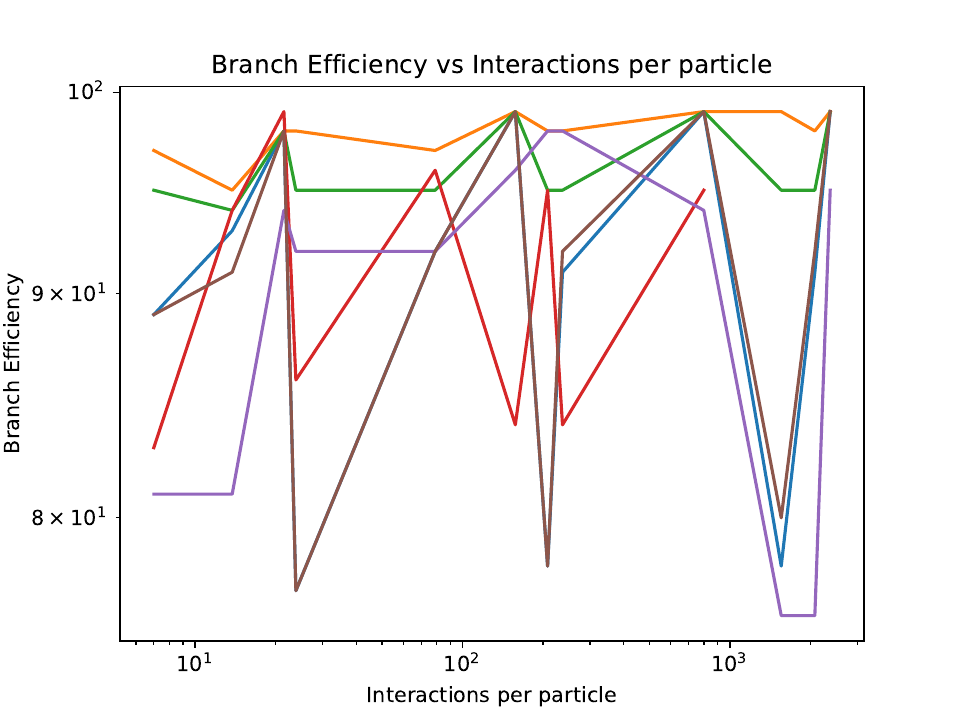}
        \caption{Branch efficiency (\%).}
        \label{fig:res:outputbranchefficiency}
    \end{subfigure}
    \hfill
    \begin{subfigure}[b]{0.48\textwidth}
        \centering
        \includegraphics[width=\textwidth, height=0.2\textheight, keepaspectratio]{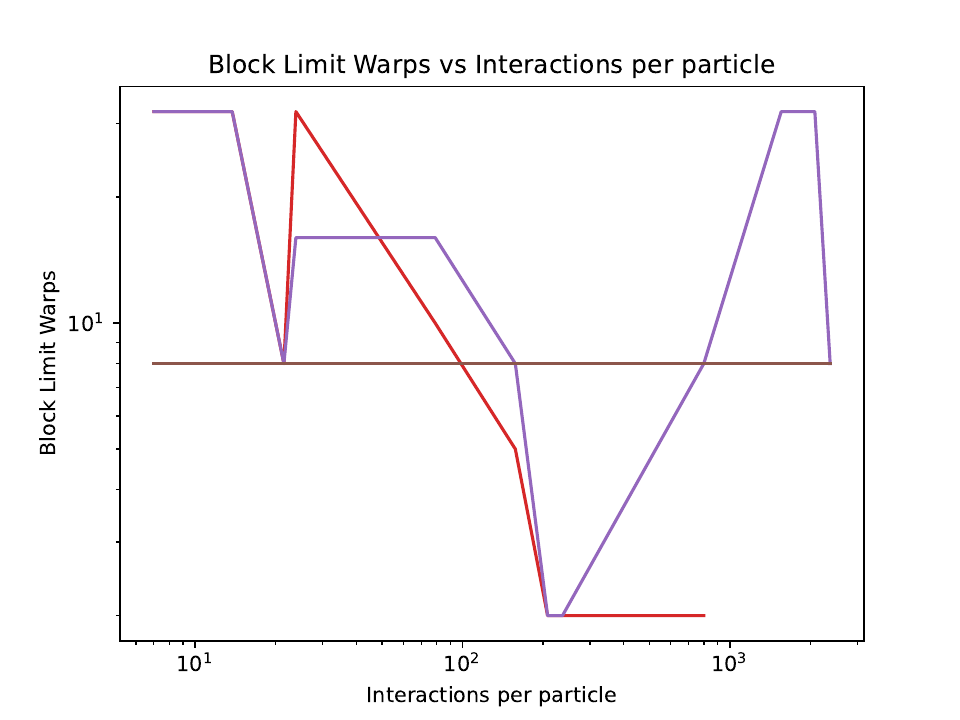}
        \caption{Block limit warps (block).}
        \label{fig:res:outputblocklimitwarps}
    \end{subfigure}

    \begin{subfigure}[b]{0.48\textwidth}
        \centering
        \includegraphics[width=\textwidth, height=0.2\textheight, keepaspectratio]{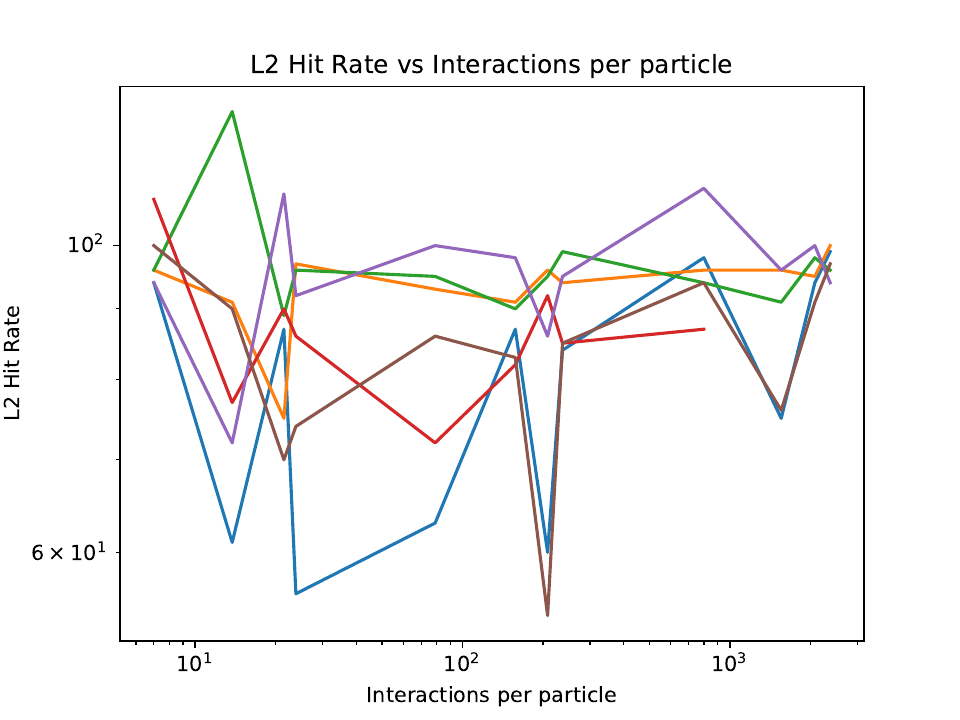}
        \caption{L2 hit rate (\%).}
        \label{fig:res:outputl2hitrate}
    \end{subfigure}
    \hfill  
    \begin{subfigure}[b]{0.48\textwidth}
        \centering
        \includegraphics[width=\textwidth, height=0.2\textheight, keepaspectratio]{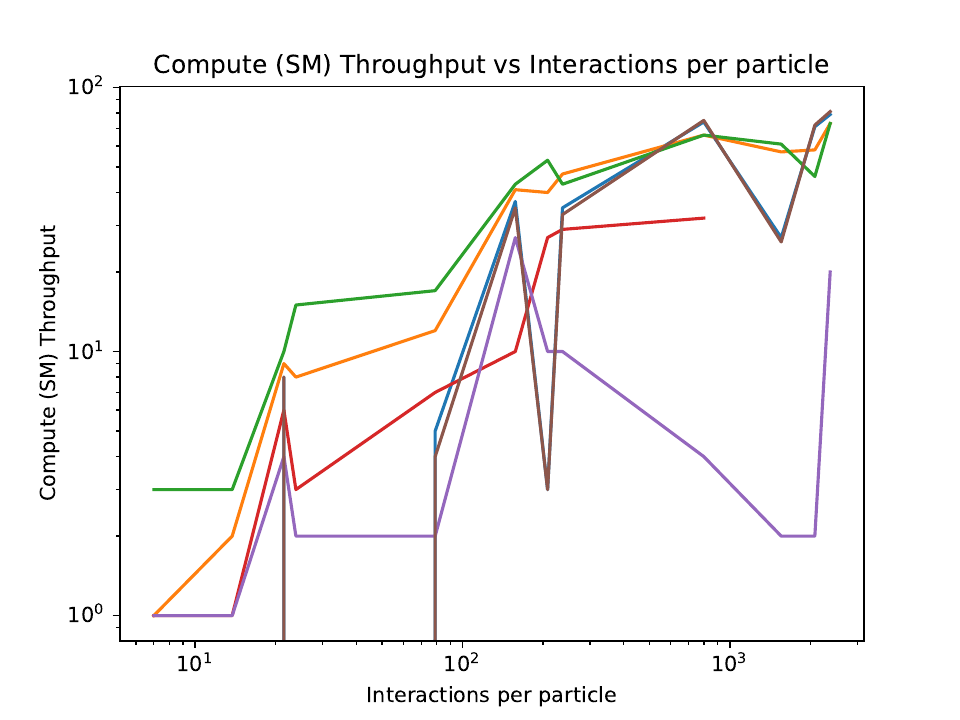}
        \caption{Compute (SM) throughput(\%).}
        \label{fig:res:outputcomputesmthroughput}
    \end{subfigure}

    \caption{Profiling results on the T600.}
    \label{fig:profs}
\end{figure}

We provide details on the profiling results using NCU (Nsight Compute CLI) on the T600 in Figure~\ref{fig:profs}.
Looking at the theoretical occupancy (Figure~\ref{fig:res:theoreticaloccupancy}), 
we can see that only our two new approaches do not always reach 100\%.
This is expected as our approaches use a large amount of shared memory, constraining the number of threads per block that can run concurrently.
The X-pencil, which uses less memory, has a 100\% potential occupancy for some configurations.
However, looking at the achieved occupancy, we can see that most implementations are far from 100\%, especially for small test cases.
While the Par-Cell-Sm has the best occupancy, it was not providing the best execution times.

Looking at the branch efficiency (Figure~\ref{fig:res:outputbranchefficiency}), we can see that most implementation have highly variable results,
except for Par-Cell and Par-Cell-Sm.
This is surprising, as these two implementations use one thread-block per cell, and then one thread per particle.
It is clear that several threads of the last warps will become quickly idle because the number of particles per cell
is not a multiple of the warp size.
However, what is also clear, is that these kernels have lots of loops (hence branches) such that since significant parts
of these loops will be the same for most of the threads in a block, the deviation is finally low.
Whereas, for Part-Par-NoLoop, there is a single branch to know if a thread id is less than the number of particles.
This test can provoke a divergence of a significant number of threads 
(up to 127 threads in our case, which is the number of threads per block minus one).
But it finally has a negligible impact on the performance.

Looking at the L2 hit ratio (Figure~\ref{fig:res:outputl2hitrate}), we can see that the strategies that are parallel over the particles
use the L2 cache less efficiently.
This is expected as the memory accesses perform by the threads can be quite different since threads of the same warps can work on particles
of different cells.
This is also valid for the All-in-SM.

Finally, in Figure~\ref{fig:res:outputcomputesmthroughput}, we provide the computed throughput (CT).
We can see that Par-cell has a better CT than Par-Cell-Sm, even if they both have similar performance.
This induces that the gain in memory access is mitigated by a penalty in the computation pipeline.
The same happens for X-pencil, which has a very low CT, even for cases where it has the better performance.

The potential gain of our two methods will drop as the arithmetic intensity increase.
Indeed, the more computation has to be made, the less memory bound will be the kernel, leading up to making the transfer from global memory negligible.
In addition, the more floating point values are needed for the computation, the fewer particles/cells will fit in shared memory, constraining even more
our approaches.

\begin{table}[htb]
\centering
\begin{tabular}{|c|c|c|c|c|c|c|}
\hline
Interactions & \multicolumn{2}{c|}{T600} & \multicolumn{2}{c|}{A100} & \multicolumn{2}{c|}{MI210} \\
per Particle & PPNL & X-pencil & PPNL & X-pencil & PPNL & X-pencil \\
\hline
\hline
7 (2/1) & 7.9e-06 & 8.3e-06 & 1.0e-05 & 1.1e-05 & 8.6e-06 & \textbf{6.9e-06} \\ \hline
13.7 (4/1) & 1.4e-05 & \textbf{9.1e-06} & 2.0e-05 & 1.2e-05 & 2.9e-05 & 1.9e-05 \\ \hline
21.4 (8/1) & 2.0e-05 & \textbf{1.2e-05} & 2.8e-05 & 1.9e-05 & 4.3e-05 & 4.1e-05 \\ \hline
23.8 (16/1) & 3.2e-05 & 3.4e-05 & 3.7e-05 & \textbf{2.0e-05} & 1.4e-04 & 9.8e-05 \\ \hline
25.3 (32/1) & 1.1e-04 & 2.3e-04 & 4.2e-05 & \textbf{3.8e-05} & 6.3e-04 & 2.3e-04 \\ \hline
79 (2/10) & 3.0e-05 & 2.6e-05 & 3.4e-05 & \textbf{1.8e-05} & 3.2e-05 & 4.2e-05 \\ \hline
157.6 (4/10) & 5.1e-05 & \textbf{3.5e-05} & 7.2e-05 & 4.6e-05 & 1.4e-04 & 7.8e-05 \\ \hline
208.1 (8/10) & 7.3e-05 & 5.7e-05 & 9.1e-05 & \textbf{5.6e-05} & 4.6e-04 & 4.6e-04 \\ \hline
236.6 (16/10) & 3.7e-04 & 4.3e-04 & 1.0e-04 & \textbf{8.1e-05} & 2.2e-03 & 1.8e-04 \\ \hline
253.2 (32/10) & 3.0e-03 & 3.8e-03 & 4.2e-04 & \textbf{4.0e-04} & 9.2e-04 & 1.2e-03 \\ \hline
799 (2/100) & 2.6e-04 & \textbf{8.4e-05} & 1.6e-04 & 1.2e-04 & 4.2e-04 & 2.7e-04 \\ \hline
1556.6 (4/100) & 4.4e-04 & \textbf{3.5e-04} & 5.2e-04 & 3.7e-04 & 3.9e-03 & 2.1e-03 \\ \hline
2079.5 (8/100) & 2.6e-03 & 3.0e-03 & 6.1e-04 & \textbf{6.1e-04} & 7.8e-03 & 5.3e-03 \\ \hline
2373.2 (16/100) & 2.2e-02 & 2.7e-02 & \textbf{3.2e-03} & 3.2e-03 & 1.2e-02 & 8.0e-03 \\ \hline
2534 (32/100) & 1.9e-01 & 2.2e-01 & 2.5e-02 & \textbf{2.5e-02} & 5.1e-02 & 5.3e-02 \\ \hline
\hline
\end{tabular}
\caption{Execution time comparison for Par-Part-NoLoop (PPNL) and X-pencil on the three hardware configurations.
The first column also includes the box division and the average number of particles per cell in parentheses.}
\label{fig:compare}
\end{table}

In Figure~\ref{fig:compare}, we provide a comparison of the different architectures, and we can notice that for a few interactions,
the differences are not significant between T600 and A100.
The MI210 seems competitive for few interactions, but then, it appears two times slower than A100.
For the larger test case, the order of the different architectures appears natural, with the A100 being the fastest, followed by the MI210 and the T600.
We remind that the MI210 has the largest shared memory capacity, but this does not provide an asset even for the All-in-SM strategy.
Also, M210 has a peak performance of 22.6TFLOP/s in single precision, while the A100 has a peak performance of 19.5TFLOP/s but remains faster.

\begin{figure}
    \centering
    \includegraphics[width=0.9\textwidth, height=\textheight, keepaspectratio]{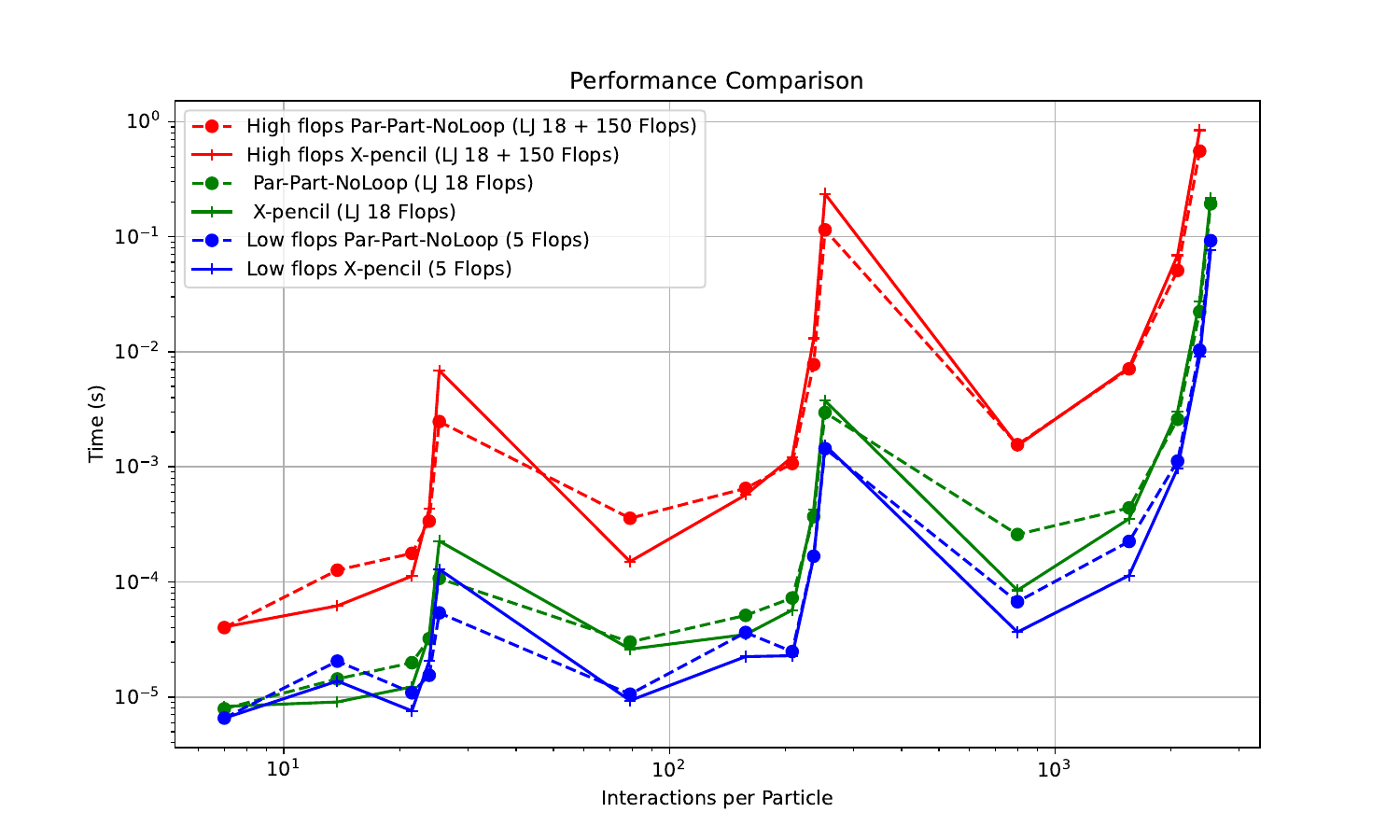}
    \caption{Comparison of the execution time for different kernel configurations on the T600.
             The Lennard-Johns kernel costs 18 FLOP per interaction.
             The low-FLOP version is obtained with a fake kernel that costs 5 FLOP per interaction (by summing the positions).
             The high-FLOP version is obtained with a fake kernel that costs 168 FLOP per interaction (it consists of the Lennard-Johns kernel with 150 added FLOP).}
    \label{fig:diffflops}
\end{figure}

In Figure~\ref{fig:diffflops}, we provide a comparison of the execution time for different kernel configurations on the T600.
We can see that as expected, the X-pencil approach is more competitive when there are few FLOP per interaction, as it make the execution more sensitive to the memory transfers.
Concerning the high-FLOP version, we can see that the X-pencil is not competitive, as the memory transfers are negligible compared to the computation.

%%%%%%%%%%%%%%%%%%%%%%%%%%%%%%%%%%%%%%%%%%%%%%%%%%%%%%%%%%%%%%%%%%%%%%%%%%%%
%%%%%%%%%%%%%%%%%%%%%%%%%%%%%%%%%%%%%%%%%%%%%%%%%%%%%%%%%%%%%%%%%%%%%%%%%%%%
\section{Related Work}
\label{sec:related-work}

% P2P
The main work on particle interactions on GPU has been proposed by Nyland et al.~\cite{nguyen2007gpu31}, to compute the gravitational potential.
They described an efficient implementation using shared memory, similar to the Par-Cell-SM strategy in our article.

% SPh
% As an example of test case where there are few particles per cell, and the use of a cutoff radius, we can cite the SPH method.
% In this method, the particles have a interaction of this type, etc.
Smooth Particles Hydrodynamics (SPH) has been used since the end of the seventies to solve astrophysical problems (\cite{gingoldSmoothedParticleHydrodynamics1977}).
More recently, it is massively used in incompressible fluids problems (\cite{koschierSurveySPHMethods2022}). 
SPH methods discretize spatial quantities by using a kernel with a given smoothing length on a set of particles. The computation of these quantities is done by interpolating with the kernel for each particle within a certain radius. More concretely, it solves the Navier Stokes equations by approximating the density of a fluid and the different forces that act on the fluid for each particle. 
% SPH methods discretizes spatial quantities using kernel with a given smoothing length on a set of particles. The computation of these quantities is made by interpolating with the kernel on each neighbors of a particles in a given radius. 
%SPH methods are specifically used to simulate incompressible flow (TODO change this formulation). 

One of the main advantage of SPH methods over classical finite element or finite volume method is the mesh free approach, for example to simulate chaotic free surface with complex boundaries. Therefore, to treat large volume of fluids, with a complexity growing in $O(n^3)$ the grid based approach is unavoidable. %Maybe add a brief line to remind the this allow us to performs the computation of a given particle $i$ only by taking the particles of the cells containing the particle $i$ and the one from the 26 neighbors cells in a range of the ssmoothing length.
%TODO add a sentence to explain that this number is related to the kernel that is chosen.
However, the SPH methods have the specificity to use particles with a low number of neighbors. For example, the cubic spline kernel which is a typical choice in this domain will have approximately $30$ to $40$ neighbors per particles \cite{ihmsenImplicitIncompressibleSPH2014},
which is what we target in this study.

There have been numerous implementations of SPH on GPU~\cite{VALDEZBALDERAS20131483,  YANG2023114514,JO2022104125,DOMINGUEZ2013617,XIA201628}. However, we did not find any studies that utilize shared memory in the manner we aim to in this paper.

%%%%%%%%%%%%%%%%%%%%%%%%%%%%%%%%%%%%%%%%%%%%%%%%%%%%%%%%%%%%%%%%%%%%%%%%%%%%
%%%%%%%%%%%%%%%%%%%%%%%%%%%%%%%%%%%%%%%%%%%%%%%%%%%%%%%%%%%%%%%%%%%%%%%%%%%%
\section{Conclusion}
\label{sec:conclusion}

In this paper, we have surveyed several methods to compute interactions between particles using a 
cutoff radius. We have introduced two new approaches that aim to utilize more shared memory. 
As expected, while the All-In-SM approach theoretically allows for the best reuse of shared memory, 
it consumes too much memory to be efficient in practice. The X-pencil approach strikes a good balance 
and yields favorable results with significant speedup in certain scenarios. 
Additionally, we have presented a straightforward implementation of prefix sum on GPU with minimal memory accesses.

The benefits of our approach could enhance performance in scenarios with lower arithmetic intensity 
or on architectures featuring more shared memory. 
However, our approach may perform poorly as arithmetic intensity increases (shifting from being memory-bound to compute-bound), leading to negligible memory access gains.
Moreover, knowing which strategy will be the fastest appears difficult to predict and require a profiling step for each kernel/configuration.
In the near future, 
we plan to assess our approach on Intel GPU, which will likely require porting our code to SYCL.

%%%%%%%%%%%%%%%%%%%%%%%%%%%%%%%%%%%%%%%%%%%%%%%%%%%%%%%%%%%%%%%%%%%%%%%%%%%%
%%%%%%%%%%%%%%%%%%%%%%%%%%%%%%%%%%%%%%%%%%%%%%%%%%%%%%%%%%%%%%%%%%%%%%%%%%%%
\section*{Acknowledgments}

Experiments presented in this paper were carried out using the PlaFRIM experimental test-bed, supported by Inria, 
CNRS (LABRI and IMB), Université de Bordeaux, Bordeaux INP and Conseil Régional d’Aquitaine (see https://www.plafrim.fr). 
This work has been partially supported by the \href{https://www.anrt.asso.fr/fr}{Association nationale de recherche et de technologie (ANRT)}
\footnote{Contract number 2021/1157} and by the \href{https://www.magelis.org/}{Pôle Image Magélis}.

%%%%%%%%%%%%%%%%%%%%%%%%%%%%%%%%%%%%%%%%%%%%%%%%%%%%%%%%%%%%%%%%%%%%%%%%%%%%
%%%%%%%%%%%%%%%%%%%%%%%%%%%%%%%%%%%%%%%%%%%%%%%%%%%%%%%%%%%%%%%%%%%%%%%%%%%%
%\bibliographystyle{plain}
%\bibliography{main}

%%%%%%%%%%%%%%%%%%%%%%%%%%%%%%%%%%%%%%%%%%%%%%%%%%%%%%%%%%%%%%%%%%%%%%%%%%%%
%%%%%%%%%%%%%%%%%%%%%%%%%%%%%%%%%%%%%%%%%%%%%%%%%%%%%%%%%%%%%%%%%%%%%%%%%%%%

\section*{Appendix}

\begin{lstlisting}[language=C++, caption={Prefix sum implementation on shared memory.}, label={lst:prefix-sum}]
template <class ElementType, typename IndexType>
__device__ void BuildPrefix_SM_device(ElementType* prefix, IndexType N) {
    // Upward pass
    IndexType js = 2;
    while (js <= N) {
        IndexType jsd2 = js / 2;
        for (IndexType idN = threadIdx.x*js + js - 1; idN < N; idN += blockDim.x*js) {
            prefix[idN] += prefix[idN - jsd2];
        }
        js *= 2;
        __syncthreads();
    }

    // Downward pass
    js = max(4, js/2);
    while (js > 1) {
        IndexType jsd2 = js / 2;
        for (IndexType idN = threadIdx.x*js + js + jsd2 - 1; idN < N; idN += blockDim.x*js) {
            prefix[idN] += prefix[idN - jsd2];
        }
        js = jsd2;
        __syncthreads();
    }
}
\end{lstlisting}

\end{document}

%% file: tikzPicture/GridExample.tikz
\tikzset{every picture/.style={line width=0.75pt}} %set default line width to 0.75pt        

\begin{tikzpicture}[x=0.75pt,y=0.75pt,yscale=-1,xscale=1]

%Shape: Grid [id:dp014458368448183045] 
\draw  [draw opacity=0] (126,87.67) -- (351.09,87.67) -- (351.09,237.87) -- (126,237.87) -- cycle ; \draw   (126,87.67) -- (126,237.87)(201,87.67) -- (201,237.87)(276,87.67) -- (276,237.87)(351,87.67) -- (351,237.87) ; \draw   (126,87.67) -- (351.09,87.67)(126,162.67) -- (351.09,162.67)(126,237.67) -- (351.09,237.67) ; \draw    ;
%Shape: Circle [id:dp32516924294867566] 
\draw  [fill={rgb, 255:red, 74; green, 144; blue, 226 }  ,fill opacity=1 ] (174.35,135.18) .. controls (174.35,130.73) and (177.96,127.12) .. (182.41,127.12) .. controls (186.87,127.12) and (190.48,130.73) .. (190.48,135.18) .. controls (190.48,139.64) and (186.87,143.25) .. (182.41,143.25) .. controls (177.96,143.25) and (174.35,139.64) .. (174.35,135.18) -- cycle ;
%Shape: Circle [id:dp8417917402258551] 
\draw  [fill={rgb, 255:red, 74; green, 144; blue, 226 }  ,fill opacity=1 ] (141.35,111.68) .. controls (141.35,107.23) and (144.96,103.62) .. (149.41,103.62) .. controls (153.87,103.62) and (157.48,107.23) .. (157.48,111.68) .. controls (157.48,116.14) and (153.87,119.75) .. (149.41,119.75) .. controls (144.96,119.75) and (141.35,116.14) .. (141.35,111.68) -- cycle ;
%Shape: Circle [id:dp6532689338359058] 
\draw  [fill={rgb, 255:red, 184; green, 233; blue, 134 }  ,fill opacity=1 ] (231.35,125.68) .. controls (231.35,121.23) and (234.96,117.62) .. (239.41,117.62) .. controls (243.87,117.62) and (247.48,121.23) .. (247.48,125.68) .. controls (247.48,130.14) and (243.87,133.75) .. (239.41,133.75) .. controls (234.96,133.75) and (231.35,130.14) .. (231.35,125.68) -- cycle ;
%Shape: Circle [id:dp10799987096593078] 
\draw  [fill={rgb, 255:red, 208; green, 2; blue, 27 }  ,fill opacity=1 ] (286.85,112.18) .. controls (286.85,107.73) and (290.46,104.12) .. (294.91,104.12) .. controls (299.37,104.12) and (302.98,107.73) .. (302.98,112.18) .. controls (302.98,116.64) and (299.37,120.25) .. (294.91,120.25) .. controls (290.46,120.25) and (286.85,116.64) .. (286.85,112.18) -- cycle ;
%Shape: Circle [id:dp1734762984192615] 
\draw  [fill={rgb, 255:red, 208; green, 2; blue, 27 }  ,fill opacity=1 ] (321.35,115.68) .. controls (321.35,111.23) and (324.96,107.62) .. (329.41,107.62) .. controls (333.87,107.62) and (337.48,111.23) .. (337.48,115.68) .. controls (337.48,120.14) and (333.87,123.75) .. (329.41,123.75) .. controls (324.96,123.75) and (321.35,120.14) .. (321.35,115.68) -- cycle ;
%Shape: Circle [id:dp9790342730069419] 
\draw  [fill={rgb, 255:red, 208; green, 2; blue, 27 }  ,fill opacity=1 ] (308.35,140.18) .. controls (308.35,135.73) and (311.96,132.12) .. (316.41,132.12) .. controls (320.87,132.12) and (324.48,135.73) .. (324.48,140.18) .. controls (324.48,144.64) and (320.87,148.25) .. (316.41,148.25) .. controls (311.96,148.25) and (308.35,144.64) .. (308.35,140.18) -- cycle ;
%Shape: Circle [id:dp2795915775470048] 
\draw  [fill=magenta  ,fill opacity=1 ] (302.35,198.68) .. controls (302.35,194.23) and (305.96,190.62) .. (310.41,190.62) .. controls (314.87,190.62) and (318.48,194.23) .. (318.48,198.68) .. controls (318.48,203.14) and (314.87,206.75) .. (310.41,206.75) .. controls (305.96,206.75) and (302.35,203.14) .. (302.35,198.68) -- cycle ;
%Shape: Circle [id:dp4301915978241767] 
\draw  [fill={rgb, 255:red, 245; green, 166; blue, 35 }  ,fill opacity=1 ] (242.85,215.18) .. controls (242.85,210.73) and (246.46,207.12) .. (250.91,207.12) .. controls (255.37,207.12) and (258.98,210.73) .. (258.98,215.18) .. controls (258.98,219.64) and (255.37,223.25) .. (250.91,223.25) .. controls (246.46,223.25) and (242.85,219.64) .. (242.85,215.18) -- cycle ;
%Shape: Circle [id:dp0014104127541600953] 
\draw  [fill={rgb, 255:red, 245; green, 166; blue, 35 }  ,fill opacity=1 ] (222.85,183.18) .. controls (222.85,178.73) and (226.46,175.12) .. (230.91,175.12) .. controls (235.37,175.12) and (238.98,178.73) .. (238.98,183.18) .. controls (238.98,187.64) and (235.37,191.25) .. (230.91,191.25) .. controls (226.46,191.25) and (222.85,187.64) .. (222.85,183.18) -- cycle ;

% Text Node
\draw (500.17,112.93) node   [align=left] {\begin{minipage}[lt]{200.97pt}\setlength\topsep{0pt}
\#Particles per cells $[\colorbox{blue!30}{2}, \colorbox{green!30}{1} ,\colorbox{red!30}{3},0,\colorbox{orange!30}{2},\colorbox{magenta!30}{1}]$\\\#Offset cell $\displaystyle [\colorbox{blue!30}{0}, \colorbox{blue!30}{2}, \colorbox{green!30}{3}, \colorbox{red!30}{6}, \colorbox{red!30}{6}, \colorbox{yellow!30}{8}, \colorbox{magenta!30}{9}]$
\end{minipage}};

\end{tikzpicture}